\documentclass[osajnl2,twocolumn,superscriptaddress]{revtex4}  
\usepackage[draft]{hyperref} 
\usepackage{amsfonts}
\usepackage{graphicx}        

\newcommand\eqnref[1]{(\ref{#1})}
\newcommand\figref[1]{Fig.~\ref{#1}}

\newcommand{\iu}   {\mathrm{i}}     
\newcommand{\Deltarm}   {\mathrm{\Delta}}

\begin{document}

\title{Superfocusing by Nano-Shells}

\author{Igor Tsukerman}
\affiliation{Department of Electrical and Computer Engineering, The
University of Akron, OH 44325-3904, USA}

\email{igor@uakron.edu}

\begin{abstract}
Recently Merlin and co-workers demonstrated, both theoretically and
in the microwave range experimentally, subwavelength focusing of
\emph{evanescent} waves by patterned plates. The present paper
extends these ideas and the design procedure to scatterers of
arbitrary shapes and to the optical range of wavelengths. The
analytical study is supported by numerical results. The most
intriguing feature of the proposed design is that, in the framework
of classical electrodynamics of continuous media, focusing can in
principle be arbitrarily sharp, subject to the constraints of
fabrication.

OCIS numbers: 310.6628, 160.4236, 350.4238, 240.6680, 250.5403,
050.1970.
\end{abstract}


\maketitle

Subwavelength focusing that circumvents the usual diffraction limit
in optics has been a very active area of research, with a multitude
of approaches explored in the literature: negative-index lenses and
guides, plasmonic particles and cascades, superoscillations,
time-reversal techniques and others
(\cite{Zheludev-What-diff-limit-08}--\cite{Dai08} and references
therein). Recently the Merlin and Grbic groups
\cite{Merlin-Radiationless-interference-07,Grbic-Near-field-plates-08}
(see also \cite{Helseth-almost-perfect-lens-08}) showed, both
theoretically and in the microwave range experimentally, that
patterned (grating-like) plates produce subwavelength focusing of
\emph{evanescent} waves if the pattern contains significantly
different spatial scales of variation. Conceptually, these patterns
are related to surface profiles of near-field optical holography
\cite{Bozhevolnyi-near-field-holography96}.

This Letter shows that the ideas of near-field focusing can be
extended to patterned nano-shells and, further, to scatterers of
arbitrary shape. The most intriguing feature is ``superfocusing'' --
i.e. focusing that can in principle be arbitrarily sharp and strong,
subject to the constraints of fabrication and availability of
materials with desired values of the dielectric permittivity
$\epsilon$. (It is also tacitly assumed that the size of the system
is sufficiently large for electrodynamics of continuous media to be
applicable; see e.g. \cite{Dai08} and references therein.)

Let us suppose that an incident plane wave with a frequency $\omega$
impinges on a scatterer coated with a plasmonic and/or dielectric
layer. In general, the shape of the scatterer may vary; more
importantly, the thickness and dielectric permittivity of the
coating may be chosen judiciously, with the ultimate goal of
nano-focusing the wave at a given spot. In practical applications,
the role of the scatterer can be played e.g. by the apex of an
optical tip, the active part of an optical sensor or by a
nano-antenna.

\begin{figure}
  \centering
  \includegraphics[width=0.8\linewidth]{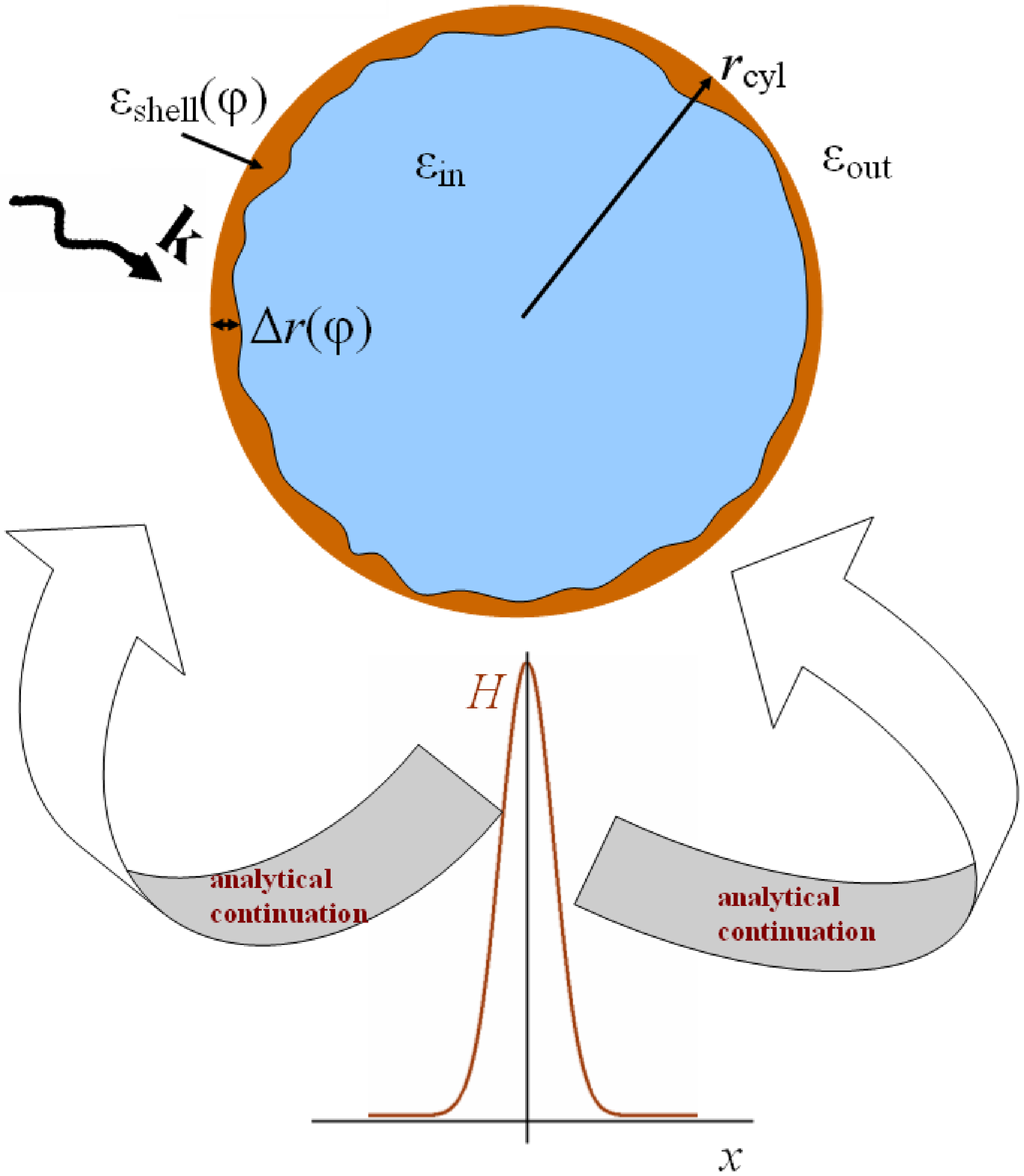}\\
  \caption{A thin shell (coating) can be capable of focusing light to an arbitrarily narrow spot
  if the angular variation of its dielectric function and/or thickness are judiciously
  chosen. See text for further details.}
  \label{fig:coating}
\end{figure}

Even though the ideas are general and could be applied in 3D
electrodynamic analysis and design, let us consider for maximum
simplicity the 2D case of a cylindrical scatterer (particle) of
radius $r_\mathrm{cyl}$ (\figref{fig:coating}). The usual complex
phasor convention with the $\exp(-\iu \omega t)$ factor is adopted.

Several ways of formulating this electromagnetic field problem are
available. First, one needs to distinguish full electrodynamic
analysis vs. quasi-static (QS) approximations valid for dimensions
much smaller than the wavelength. The QS treatment has limitations,
as it does not account for wave effects and short-range surface
plasmon modes \cite{Valle-Sondergaard-Bozhevolnyi-SPP-resonators08}.
However, to fix ideas and for the sake of algebraic simplicity this
Letter adopts the QS approximation. Full wave analysis is analogous
but involves Bessel/Hankel functions and their derivatives instead
of polynomials in $r$ (see expressions below).

Second, the electrodynamic problem may be formulated in terms of the
electric or magnetic field (convenient for $s$- and
$p$-polarizations, respectively). Alternatively, two dual
formulations are also available under the QS approximation: via the
electric scalar potential and via the stream function
$\mathbf{\Psi}$, $\nabla \times \mathbf{\Psi} = \epsilon \mathbf{E}$
\footnote{The magnetic field can itself be viewed as a (scaled)
stream function.}; see \cite{Shvets-Urzhumov-04,Mayergoyz05} for
some interesting implications of this duality.

Let us endeavor to achieve sharp focusing with respect to the polar
angle $\phi$ at a certain radius $r_f > r_\mathrm{cyl}$. For
$p$-polarization (the two-component electric field in the
cross-sectional plane and the one-component magnetic field $H$
directed along the axis of the particle), let
\begin{equation}\label{eqn:desired-H-at-rf}
    H_\mathrm{out} (r_f, \phi) ~=~ H_f g(\phi - \phi_f)
\end{equation}
where $H_\mathrm{out}$ is the magnetic field outside the shell,
$H_f$ is some amplitude at the focus, $\phi_f$ is a given angle and
$g(\phi)$ is a desired distribution of the field. For example,
choosing $g(\phi)$ or, alternatively, $g'(\phi)$ as a sharp Gaussian
peak will result in the corresponding peak in the magnetic field or
the radial component of the electric field, respectively.
%

This local behavior of the field can be extended, by analytical
continuation, to the whole region outside the scattering shell. One
way of doing so is by expanding the field into cylindrical harmonics
whose coefficients are found from the Fourier transform of $g$. This
analytical continuation will then determine the field distribution
on the surface of the particle, and one needs to find the parameters
of the coating that would produce such a surface distribution.

%
%
More specifically, in the QS limit the field inside the shell can be
expanded into cylindrical harmonics:
%
\begin{equation}\label{eqn:cyl-polynomial-harmonics-in}
    H_\mathrm{in}(r, \phi) ~=~  \sum_{n=-\infty}^{\infty} a_n r^{|n|} \exp(\iu n \phi) ,
    ~~ r \le r_{\mathrm{cyl}}
\end{equation}
where $a_n$ are some coefficients.
%
%
Outside of the cylinder ($r > r_{\mathrm{cyl}}$) the magnetic field
can be represented as the sum of the incident field and the
scattered field:
\begin{equation}\label{eqn:total-quasi-static-field-eq-inc-plus-scattered}
    H_\mathrm{out} ~=~ H_\mathrm{INC} \,+\, H_\mathrm{s}
\end{equation}
The incident field under the QS approximation is linear with respect
to $r$:
\begin{equation}\label{eqn:incident-field-quasi-static}
    H_\mathrm{INC} ~=~ H_0 \,+\, h_0 r \cos(\phi - \phi_0)
\end{equation}
where $h_0$ is a coefficient. The constant $H_0$ does not affect the
results and is for simplicity set to zero in the remainder. The
scattered field is
\begin{equation}\label{eqn:cyl-harmonics-quasi-static-out}
    H_\mathrm{s}(r, \phi) ~=~  \sum_{n=-\infty}^{\infty} c_n
    r^{-|n|} \exp(\iu n \phi)
\end{equation}
%
%
where $c_n$ are coefficients to be determined.

The boundary conditions across the coating are as follows. If the
coating is nonmagnetic \footnote{True for all natural materials at
optical frequencies.} and thin, the magnetic flux passing through it
can be neglected and consequently the tangential component $E_\phi$
of the electric field is assumed to be continuous across the
coating:
\begin{equation}\label{eqn:no-jump-E-across-coating-qs}
    E_{\phi, \mathrm{out}} ~=~ E_{\phi, \mathrm{in}}
\end{equation}
\vskip -0.1in \noindent Substituting
\begin{equation}\label{eqn:E-phi-eq-dH-dr}
    E_\phi ~=~ \frac{1}{\iu \omega \epsilon} \frac{\partial H}
    {\partial r}
\end{equation}
\vskip -0.1in \noindent into
\eqnref{eqn:no-jump-E-across-coating-qs}, we have
%
%
\begin{equation}\label{eqn:dH-dr-quasi-static-inside}
    \frac{dH_\mathrm{in}}{dr} ~=~
    \frac{ \epsilon_\mathrm{in}}{ \epsilon_\mathrm{out}} \,
    \frac{dH_\mathrm{out}}{dr}
\end{equation}
which links the coefficients $a_n$, $c_n$ as follows:
%
\begin{equation}\label{eqn:an-quasi-static}
    a_n ~=~ -c_n r_\mathrm{cyl}^{-2|n|} \, \frac{ \epsilon_\mathrm{in}}{ \epsilon_\mathrm{out}}
    , ~~~~ n \neq \pm 1
\end{equation}
%
\vskip -0.1in
\begin{equation}\label{eqn:a1-quasi-static}
    a_{\pm 1} ~=~
    \frac{ \epsilon_\mathrm{in}}{ \epsilon_\mathrm{out}} \, \left(
    -c_{\pm 1} r_\mathrm{cyl}^{-2} \,+\, \frac12 h_0 \exp(\mp \iu
    \phi_0) \right)
\end{equation}
The jump of the magnetic field is not neglected, as it corresponds
to a current layer that may be appreciable if
$|\epsilon_\mathrm{shell}|$ is large (e.g. for plasmonic materials):
\begin{equation}\label{eqn:jump-H-quasi-static-across-coating}
    H_\mathrm{out} - H_\mathrm{in} ~=~ \iu \omega \,
    \epsilon_\mathrm{shell}(\phi) \,  E_\phi \, \Delta r(\phi)
\end{equation}
With $E_\phi$ defined by \eqnref{eqn:E-phi-eq-dH-dr}, equation
\eqnref{eqn:jump-H-quasi-static-across-coating} becomes
\begin{equation}\label{eqn:jump-H-quasi-static-via-dH-dr}
    H_\mathrm{out} - H_\mathrm{in} ~=~
    \frac{ \epsilon_\mathrm{shell}(\phi)} {\epsilon_\mathrm{in/out}} \,
    \Delta r(\phi) \,
    \frac{\partial H_\mathrm{in/out}} {\partial r}
\end{equation}
where the ``in/out'' subscript in the right hand side encompasses
two equally valid expressions: one with $\epsilon_{\mathrm{in}}$ and
$H_{in}$ and another one with $\epsilon_{\mathrm{out}}$ and
$H_{\mathrm{out}}$.

\textbf{From superfocusing at a point to fields on the cylinder}.
Let the desired field distribution around a certain focusing point
$r = r_f > r_{\mathrm{cyl}}$, $\phi = \phi_f$ be given by
\eqnref{eqn:desired-H-at-rf}. This behavior of the field defines the
expansion coefficients $c_n$:
\begin{equation}\label{eqn:cn-quasi-static}
    c_n ~=~ \frac{H_f} {r_f^{|n|}} \, \tilde{g}_n, ~~~ n \neq \pm 1;
    ~~~ c_{\pm 1} = \frac{H_f} {r_f} - \frac12 h_0 r_f \exp(\mp \iu \phi_0)
\end{equation}
where $\tilde{g}_n = (2\pi)^{-1} \int\nolimits_0^{2\pi} g(\phi -
\phi_f) \exp(\iu n \phi) d \phi$ are the Fourier coefficients.

With the coefficients $c_n$ so defined, one can now evaluate $a_n$
from \eqnref{eqn:an-quasi-static}, \eqnref{eqn:a1-quasi-static},
then the jump of the magnetic field and consequently, from
\eqnref{eqn:jump-H-quasi-static-via-dH-dr}, the required parameters
of the coating.
\begin{equation}\label{eqn:eps-delta-r-eq-jump-H-via-E-quasi-static}
    \epsilon_\mathrm{shell}(\phi) \, \Deltarm r(\phi) ~=~
    \frac{ \epsilon_\mathrm{in} \, (H_\mathrm{out}(r_\mathrm{cyl}) - H_\mathrm{in}(r_\mathrm{cyl}))}
    {\partial H_\mathrm{in}(r_\mathrm{cyl}) / \partial r}
\end{equation}
This result is valid for thin nonmagnetic shells and also holds for
the full electrodynamic problem, although the coefficients $a_n$,
$c_n$ implicit in
\eqnref{eqn:eps-delta-r-eq-jump-H-via-E-quasi-static} are different
in the wave case.\footnote{As in \cite{Grbic-Near-field-plates-08},
in the wave case small negative values of
$\epsilon_\mathrm{shell}''(\phi)$ in
\eqnref{eqn:eps-delta-r-eq-jump-H-via-E-quasi-static} are
mathematically possible but could in practice ignored.}

Note that Eq. \eqnref{eqn:eps-delta-r-eq-jump-H-via-E-quasi-static}
defines the \emph{product} of the permittivity and thickness of the
shell rather than each of these quantities separately. This reflects
the impedance-like nature of the thin-shell boundary condition: $Z
\equiv E_\phi / (H_\mathrm{out} - H_\mathrm{in}) = \iu / (\omega
\epsilon_\mathrm{shell} \Deltarm r)$ has the physical meaning of
impedance.

%
%
%

\begin{figure}
  \centering
  \includegraphics[width=0.95\linewidth]{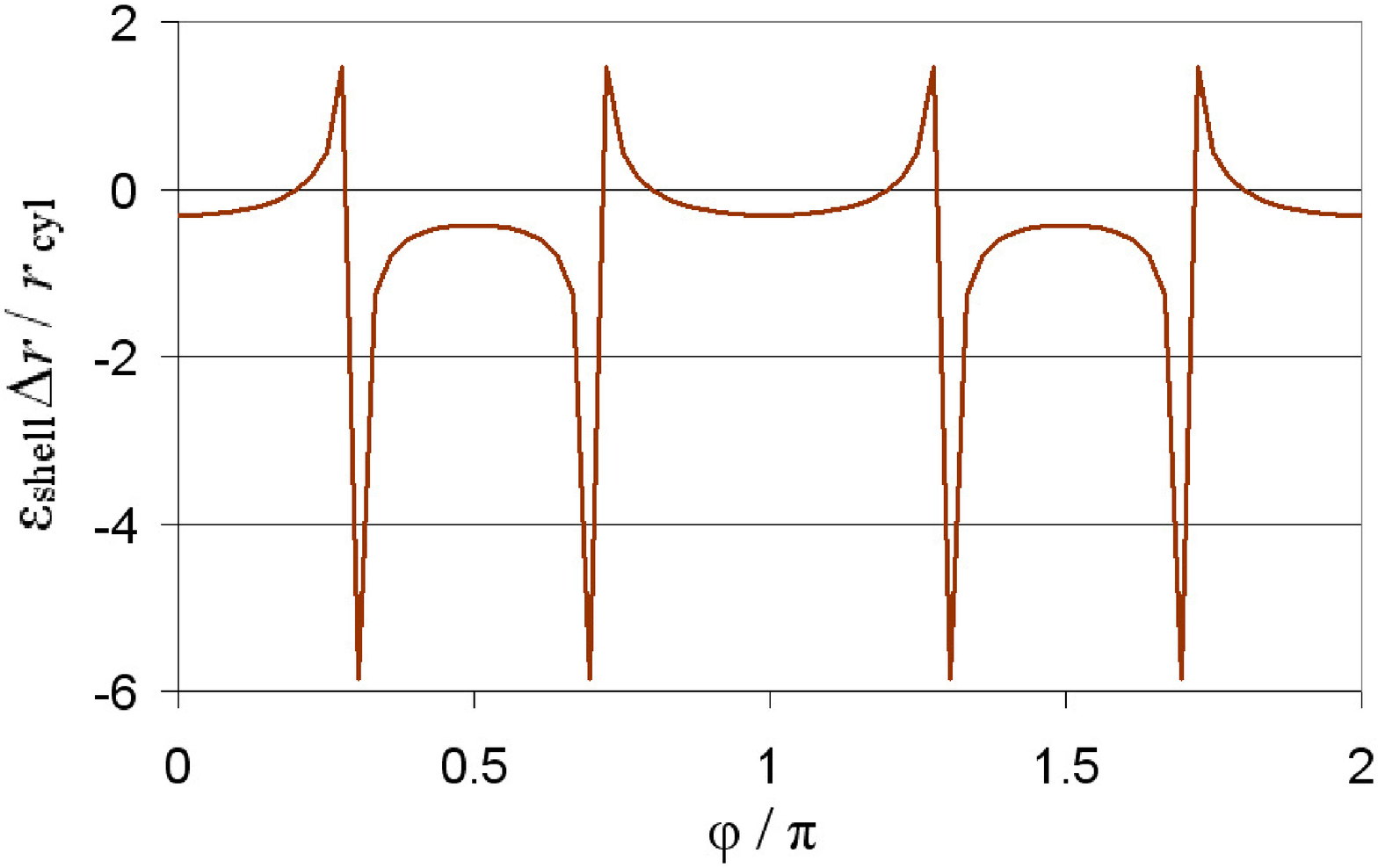}\\
  \caption{$\epsilon_\mathrm{shell} \, \Deltarm r / r_\mathrm{cyl}$ vs. angle for
  $H_f = 2$, $r_f = 1.2 r_\mathrm{cyl}$ and $\Deltarm \phi = 2\pi / 16$, $n_{\max} = 12$ harmonics.}
  \label{fig:eps-dr-vs-phi-12harm-Hf2-rf1.2-dphi-2pi16}
\end{figure}

To summarize, the algorithm proceeds as follows:
\begin{enumerate}
    \item Choose $r_\mathrm{cyl}$, $r_f$ and the maximum number $n_{\max}$ of
    cylindrical harmonics.
    \item Compute $c_n$ from \eqnref{eqn:cn-quasi-static}, for $|n|
    \leq n_{\max}$.
%
%
    \item Find $a_n$ from \eqnref{eqn:an-quasi-static},
    \eqnref{eqn:a1-quasi-static}.
    %
    \item Compute $H_\mathrm{out}(r_\mathrm{cyl})$,
    $H_\mathrm{in}(r_\mathrm{cyl})$ and ${\partial H_\mathrm{in}(r_\mathrm{cyl}) / \partial
    r}$ using the cylindrical harmonic expansion with the coefficients $a_n$, $c_n$
    found previously.
    \item Find $\epsilon_\mathrm{shell}(\phi) \, \Deltarm r (\phi)$ from
    \eqnref{eqn:eps-delta-r-eq-jump-H-via-E-quasi-static}.
\end{enumerate}
Changing the desired field variation $g(\phi)$ and the amplitude
$H_f$, one obtains a very rich variety of solutions for the shell
parameters \eqnref{eqn:eps-delta-r-eq-jump-H-via-E-quasi-static}. To
illustrate some of the possibilities, consider the angular
distribution of $\epsilon_\mathrm{shell} \, \Deltarm r /
r_\mathrm{cyl}$ at $r = r_f = 1.2 r_\mathrm{cyl}$. For relatively
small values of $H_f$, the sign of $\epsilon_\mathrm{shell}$ varies
(\figref{fig:eps-dr-vs-phi-12harm-Hf2-rf1.2-dphi-2pi16}, $H_f = 2$,
$\Deltarm \phi = 2\pi / 16$, $n_{\max} = 12$).

\begin{figure}
  \centering
  \includegraphics[width=0.95\linewidth]{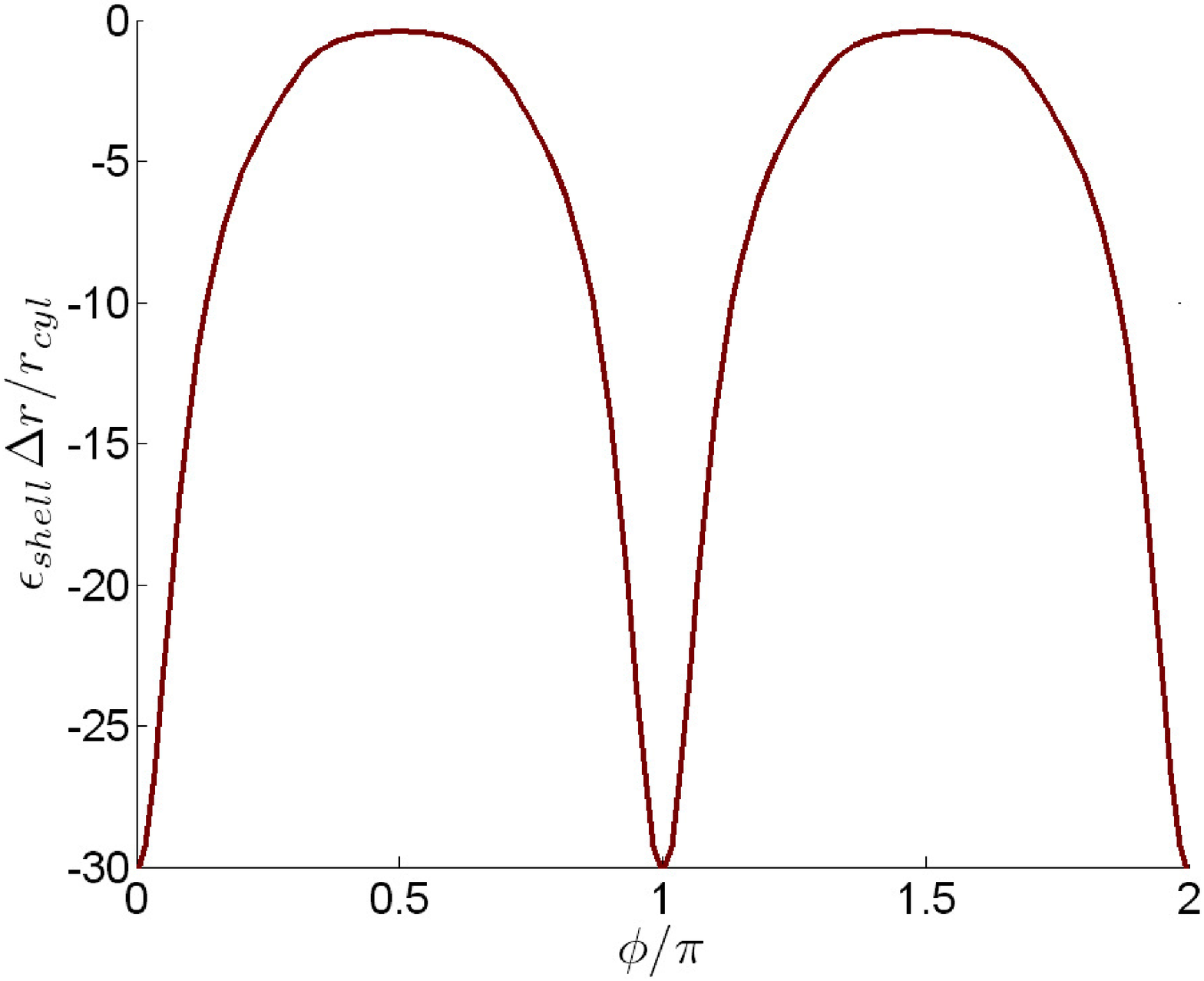}\\
  \caption{$\epsilon_\mathrm{shell} \, \Deltarm r / r_\mathrm{cyl}$ vs. angle for
  $H_f = 10$, $r_f = 1.2 r_\mathrm{cyl}$ and $\Deltarm \phi = 2\pi / 18$, $n_{\max} = 20$ harmonics.}
  \label{fig:eps-dr-vs-phi-20harm-Hf10-rf1.2-w18.eps}
\end{figure}

\begin{figure}
  \centering
  \includegraphics[width=0.95\linewidth]{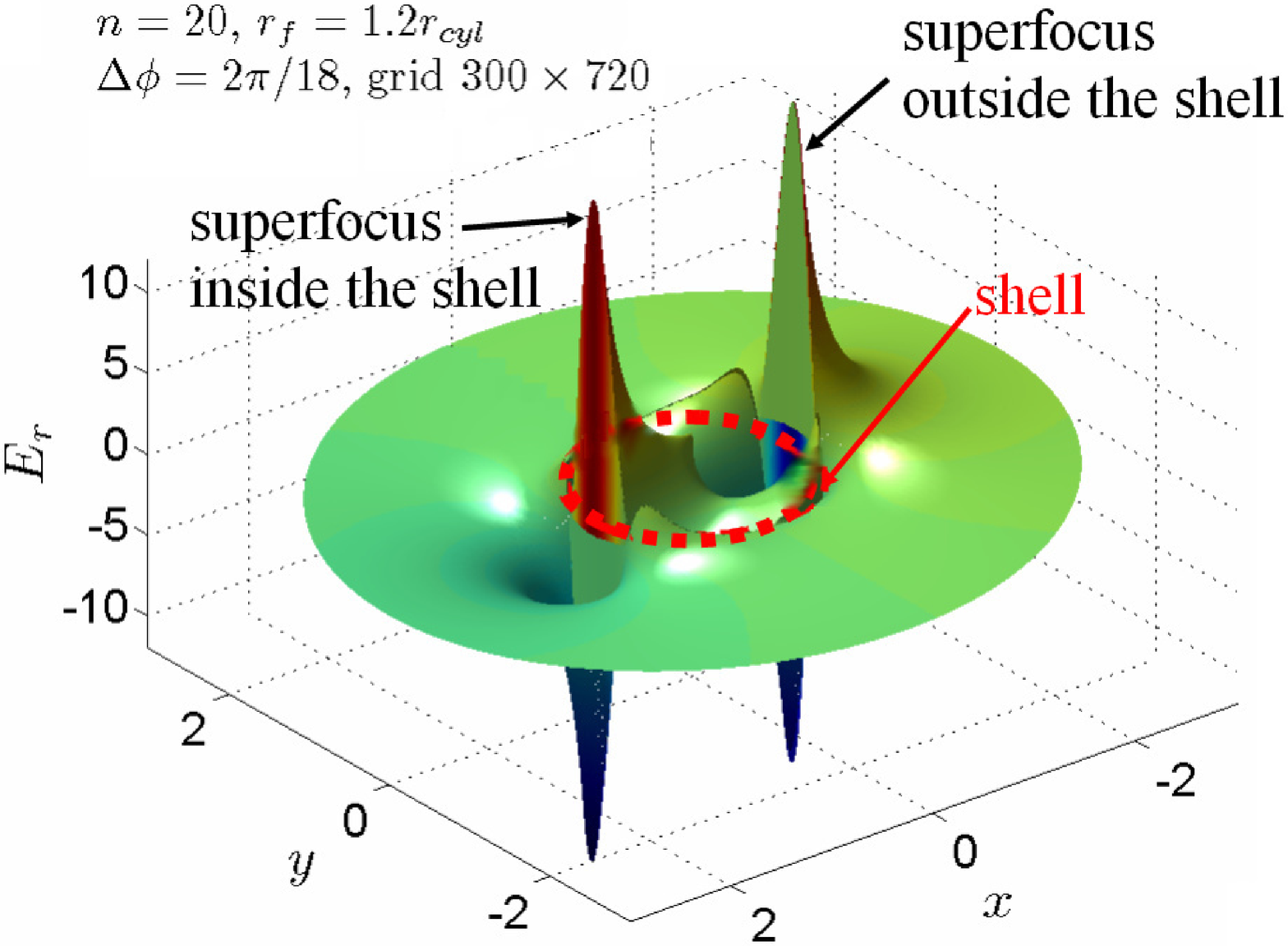}\\
  \caption{Surface plot of $E_r$ reveals ``super-focuses'' inside
  and outside the shell. No losses. Parameters: $H_f = 10$, $n_{\max} = 20$ harmonics,
  $r_f = 1.2 r_\mathrm{cyl}$, $\Deltarm \phi = 2\pi / 18$, FD grid $n_r
\times n_\phi = 300 \times 720$.}
  \label{fig:Er-surfplot-20harm-Hf10-rf1.2-dphi-2pi18-300x720}
\end{figure}

\begin{figure}
  \centering
 \includegraphics[width=0.98\linewidth]{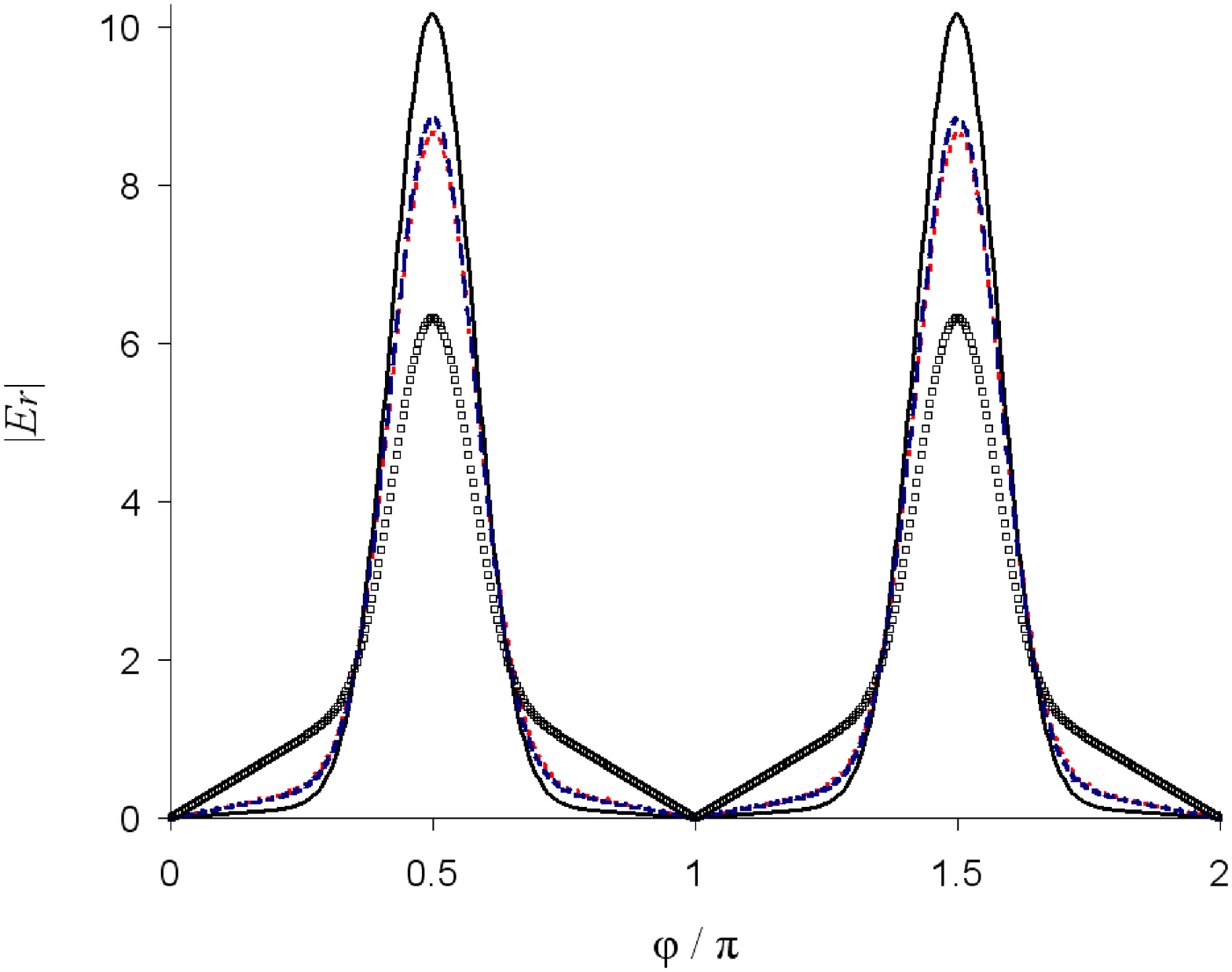}\\
  \caption{$E_r$ vs. angle at the radius of the focus $r_f = 1.2 r_\mathrm{cyl}$.
  Same parameters as in the previous figure. Losses ($\epsilon''_{\mathrm{shell}}
  = 0.1$) reduce the magnitude of the focusing peak but do not significantly affect its sharpness.
  Solid line: semi-analytical solution (20 harmonics). Dashed and
dotted lines on top of one another: no losses, FD solutions for
grids $n_r \times n_\phi = 150 \times 480$ and $300 \times 720$.
Empty squares: FD solution, $\epsilon'' = 0.1$, grid $150 \times
480$.}
  \label{fig:Er-vs-phi-n20-w18-dr-0.02}
\end{figure}

Higher values of $H_f$ require a strong plasmonic resonance with
$\epsilon_\mathrm{shell}(\phi)$ consistently negative
(\figref{fig:eps-dr-vs-phi-20harm-Hf10-rf1.2-w18.eps}, $H_f = 10$,
$n_{\max} = 20$ and $\Deltarm \phi = 2\pi / 18$). Here the
derivative $g'(\phi)$ was chosen as a Gaussian peak $g'(\phi) =
\exp(-\phi^2 / \Deltarm \phi^2)$, where parameter $\Deltarm \phi$
controls the width of the peak. This angular distribution of the
magnetic field leads to a respective peak in $E_r(\phi)$
(\figref{fig:Er-surfplot-20harm-Hf10-rf1.2-dphi-2pi18-300x720} and
\figref{fig:Er-vs-phi-n20-w18-dr-0.02}). The fields were computed
semi-analytically with
\eqnref{eqn:total-quasi-static-field-eq-inc-plus-scattered}--\eqnref{eqn:cyl-harmonics-quasi-static-out}
and also numerically using finite-difference (FD) analysis on
regular polar grids. The semi-analytical solution in its present
form is valid only in the absence of losses
($\epsilon''_\mathrm{shell} = 0$) and, as
\figref{fig:Er-vs-phi-n20-w18-dr-0.02} shows, is in that case close
to the numerical solution. (The discrepancy is due primarily to the
finite thickness of the shell $\Deltarm r = 0.02 r_\mathrm{cyl}$ in
FD simulations, and to the limited number of cylindrical harmonics
in the semi-analytical treatment.) Losses affect the amplitude of
the peak but not its sharpness
(\figref{fig:Er-vs-phi-n20-w18-dr-0.02}, $\epsilon''_\mathrm{shell}
= 0.1$). Convergence of FD results was verified by running the
simulation on different grids (e.g. almost coinciding dashed and
dotted lines corresponding to two different grids,
\figref{fig:Er-vs-phi-n20-w18-dr-0.02}).

In summary, ``superfocusing'' of light by nanoshell scatterers, with
the sharpness of the focus in principle unlimited, has been
demonstrated analytically and numerically. The shell is designed in
two stages: (i) analytical continuation of the desired behavior of
the field at the focus to the boundary of the scatterer; (ii)
finding the distribution of the dielectric permittivity and
thickness of the shell that would produce the required field on the
surface. Recently proposed focusing by ``superoscillations''
\cite{Zheludev-Huang09} also falls into this framework, but with the
waveform at the focus cleverly chosen to contain only traveling wave
components.

The procedure can be generalized to arbitrary shapes of the
scattering shell and to 3D by still using cylindrical / spherical
harmonic expansion in the outside region (as in T-matrix methods
\cite{Mishchenko-Travis-Lacis02}) and numerical methods (e.g. FEM)
inside.

Numerical examples of very sharp focusing presented in this Letter
should be viewed primarily as proof of concept; further improvement
of the focusing effects should definitely be possible with
sophisticated numerical optimization techniques (e.g. adaptive
goal-oriented finite element analysis \cite{Oden-Prudhomme-01}) that
will treat not only the physical properties of the shell but also
its geometric shape as adjustable parameters. Finally, it would be
interesting to explore similar ideas for superfocusing of surface
plasmon polaritons propagating on surfaces with judiciously chosen
patterns \cite{Zheludev-Huang09}.

I thank N. I. Zheludev, S. I. Bozhevolnyi and the anonymous reviewer
for very interesting and helpful comments.

\bibliographystyle{plain}

\vskip -0.25in
\begin{thebibliography}{99}

\bibitem{Zheludev-What-diff-limit-08}
N.I. Zheludev.
\newblock What diffraction limit?
\newblock {\em Nature Materials}, 7(6):420--422, 2008.

\bibitem{Tsakmakidis-Trapped-rainbow-07}
K.L. Tsakmakidis, A.D. Boardman \& O. Hess.
\newblock `Trapped rainbow' storage of light in metamaterials.
\newblock {\em Nature}, 450(7168):397--401, 2007.


\bibitem{Pendry-TR-NIM-08}
J. Pendry.
\newblock Time reversal and negative refraction.
\newblock {\em Science}, 322(5898):71--73, 2008.

\bibitem{Zheludev-Huang09}
N. Zheludev, F. Huang.
\newblock Superresolution without evanescent waves.
\newblock \emph{Nano Letters}, in press.
\newblock N.~I.~Zheludev, private communication, Jan 2009.


\bibitem{Pendry04}
J.B. Pendry and D.R. Smith.
\newblock Reversing light with negative refraction.
\newblock {\em Phys. Today}, 57:37--43, 2004.

\bibitem{Dai08}
J. Dai, F. \v{C}ajko, I. Tsukerman, and M.I. Stockman.
\newblock Electrodynamic effects in plasmonic nanolenses.
\newblock {\em Phys. Rev. B}, 77:115419-1-5, 2008.


\bibitem{Merlin-Radiationless-interference-07}
R. Merlin.
\newblock Radiationless electromagnetic interference: Evanescent-field lenses and perfect focusing.
\newblock {\em Science}, 317(5840):927--929, 2007.

\bibitem{Grbic-Near-field-plates-08}
A. Grbic, L. Jiang, R. Merlin.
\newblock Near-field plates: Subdiffraction focusing with patterned surfaces.
\newblock {\em Science}, 320 (5875):511--513, 2008.


\bibitem{Helseth-almost-perfect-lens-08}
L.E. Helseth.
\newblock The almost perfect lens and focusing of evanescent waves.
\newblock {\em Opt Comm}, 281(8):1981--1985, 2008.


\bibitem{Bozhevolnyi-near-field-holography96}
S. I. Bozhevolnyi and B. Vohnsen.
\newblock Near-field optical holography.
\newblock {\em Phys. Rev. Lett.}, 77:3351--3354, 1996.
\newblock S.I.~Bozhevolnyi, private communication, Jan 2009.

\bibitem{Valle-Sondergaard-Bozhevolnyi-SPP-resonators08}
G. Della Valle, T. Sondergaard, and S. I. Bozhevolnyi,
Plasmon-polariton nano-strip resonators: from visible to infra-red,
\emph{Opt. Express} 16, 6867--6876, 2008.


\bibitem{Mayergoyz05}
I. D. Mayergoyz, D. R. Fredkin and Z. Zhang, Electrostatic (plasmon)
resonances in nanoparticles, \emph{Phys. Rev. B} 72, 155412, 2005.

\bibitem{Shvets-Urzhumov-04}
G. Shvets and Y. A. Urzhumov.
\newblock Engineering the electromagnetic properties of periodic
nanostructures using electrostatic resonances.
\newblock {\em Phys. Rev. Lett.}, 24:243902-1--4, 2004.

\bibitem{Oden-Prudhomme-01}
J.T. Oden, S. Prudhomme.
\newblock Goal-oriented error estimation and adaptivity for the finite element method.
\newblock {\em Computers \& Mathematics with Applications}, 41(5--6):735--756, 2001.

\bibitem{Mishchenko-Travis-Lacis02}
M. I. Mishchenko, L. D. Travis, and A. A. Lacis, \emph{Scattering,
Absorption, and Emission of Light by Small Particles}, Cambridge
University Press, Cambridge, 2002.
\end{thebibliography}

\end{document}